\documentclass[
reprint,superscriptaddress,
amsmath,amssymb,prl,
]{revtex4-1}
\usepackage{amssymb}
\usepackage{amsmath}
\usepackage{graphicx}
\usepackage{epsfig}
\usepackage{rotating}
\usepackage{verbatim}
\usepackage{wrapfig}
\usepackage{float}
\usepackage[export]{adjustbox}
\usepackage{subcaption}
\usepackage[skip=2pt,font=small]{caption}
\usepackage{ftnxtra}
\usepackage{physics}
\usepackage{color}
\usepackage{graphicx}
\graphicspath{{figures/}}
\usepackage{dcolumn}
\usepackage{bm}

\usepackage[
total={6.5in,9.75in}, top=0.9 in, left=0.9in, includefoot,
]{geometry}
\begin{document}

\preprint{APS/123-QED}

\title{Anomalous dispersion of microswimmer populations}

\author{Laxminarsimharao Vennamneni}
\affiliation{Department of Mechanical Engineering, Gandhi Institute of Technology and Management, Hyderabad 502329, India}
\author{Piyush Garg}
\affiliation{Engineering Mechanics Unit, Jawaharlal Nehru Centre for Advanced Scientific Research,
Bangalore 560064, India}
\author{Ganesh Subramanian}
 \email{sganesh@jncasr.ac.in}
\affiliation{Engineering Mechanics Unit, Jawaharlal Nehru Centre for Advanced Scientific Research,
Bangalore 560064, India}


\begin{abstract}
We examine the longitudinal dispersion of spheroidal microswimmers in pressure-driven channel flow. When time scales corresponding to swimmer orientation relaxation, and diffusion in the gradient and flow directions, are well separated, a multiple scales analysis leads to the shear-enhanced diffusivity governing the long-time spread of the swimmer population along the flow\,(longitudinal) direction. For large $Pe_r$, $Pe_r$ being the rotary Peclet number, this diffusivity scales as $O(Pe_r^4D_t)$ for $1 \leq \kappa \lesssim 2$, and as $O(Pe_r^{\frac{10}{3}}D_t)$ for $\kappa = \infty$, $D_t$ being the (bare)\,swimmer translational diffusivity and $\kappa$ the swimmer aspect ratio. For $2 \lesssim \kappa < \infty$,  swimmers collapse onto the centerline with increasing $Pe_r$, leading to an anomalously reduced diffusivity of $O(Pe_r^{5+C(\kappa)}D_t)$. Here, $C(\kappa)\!<\!-1$ characterizes the algebraic decay of swimmer concentration outside an $O(Pe_r^{-1})$ central core, with the anomalous exponent governed by large velocity variations sampled by the few swimmers outside this core. $C(\kappa)$ dips below $-5$ for $\kappa \gtrsim 10$, leading to a flow-independent bound of $O(\kappa^{10}D_t)$ for the dispersion of sufficiently slender swimmers.
\end{abstract}

\maketitle



GI Taylor showed that the cross-sectionally averaged solute concentration in pipe Poiseuille flow, in a reference frame moving with the mean velocity, spreads diffusively in the axial direction with an effective diffusivity $D_{ef\!f}\!=\!D\!\left[\!1\!+\!\frac{Pe^2}{48}\!\right]$, where $D$ is the solute diffusivity and $Pe\!=\!U_mR/D$ is the Peclet number based on the pipe radius $R$ and the centerline speed $U_m$\cite{Taylor_1953,Aris_1956}; the analogous result for pressure-driven channel flow is $D_{ef\!f}\!=\!D\!\!\left[\!1\!+\!\frac{Pe^2}{210}\!\right]$ with $Pe\!=\!U_mH/D$, $H$ being the channel half-width\cite{Wooding_1960,Dorfman_2001}. The shear-enhanced contribution to $D_{ef\!f}$ in these cases may be obtained from the flow-aligned component of the variance, $\langle x^2 \rangle$, of a Brownian particle\,(solute molecule) in the relevant flow. For a particle starting at the channel centerline\,($z\!=\!0$, $z$ being the gradient coordinate), convection by the variable component of the flow $u\!\sim\!U_m\frac{z^2}{H^2}$ gives $x\!\sim\!(U_mt)\frac{z^2}{H^2}$, so $\langle x^2 \rangle\!\sim\!\frac{(U_mt)^2}{H^4}\langle z^4 \rangle$. Using $\langle z^4 \rangle\!\sim\!D^2t^2$ for the Gaussian distribution, that develops in the gradient direction due to diffusion alone, gives $\langle x^2 \rangle\!\sim\!\frac{(U_m D)^2}{H^4} t^4$. The super-diffusive growth changes to a diffusive one when the particle samples the maximum velocity variation of $O(U_m)$ for $z \sim O(H), t \sim O(H^2/D)$. Writing $t^4$ above as $t^3.t$, replacing $t^3$ by $H^6/D^3$, and comparing with $\langle x^2 \rangle \sim D_{ef\!f}t$, gives the $Pe$-dependent term in $D_{ef\!f}$\footnote{Unlike a linear flow\cite{Foister_1980,SubBrady_2004}, the mean position of the Brownian particle also evolves with time in a reference frame moving with $U_m$, with analogous arguments giving $<x> \sim -\frac{U_m D}{H^2} t^2$ for short times, and $<x> \sim -U_m t$ for times longer than $O(H^2/D)$, reflecting the fact that the solute distribution moves with the mean speed rather than the centerline maximum. The variance, $<\!\!x^2\!\!>-<\!\!x\!\!>^2$, scales in the same manner as $<\!\!x^2\!\!>$.}; the bare molecular diffusivity\,($D$) enters as an independent additive contribution\cite{Aris_1956}. Note that while $D_{ef\!f}$ increases quadratically with $Pe$, transition to the diffusive regime requires a progressively longer time of $O(Pe.H/U_m)$.

Herein, we examine the Taylor dispersion of a population of spheroidal microswimmers in pressure-driven channel flow. The problem has a direct bearing on the spread of microbial populations in flowing environments, and is interesting due to the range of behavior exhibited by microswimmers even in  simple unidirectional shearing flows\cite{Stark2013,Bearon_2015,Saintillan_2015,Rusconi_2014,Barry_2015,vennamneni2020shear,Ryabov_2021,Cencini_2022}. In the classical scenario above, the passive solute spreads uniformly across the cross-section for $t\!>\!O(H^2/D)$. In contrast, non-spherical swimmers exhibit either high-shear\,(near-wall)\cite{Bearon_2015,Saintillan_2015,Rusconi_2014,vennamneni2020shear} or low-shear\,(near-center)\cite{Barry_2015,vennamneni2020shear,Ryabov_2021,Cencini_2022} trapping depending on $Pe_r$ and $\kappa$\footnote{An anisotropic passive solute\,(Brownian fiber) is predicted to exhibit high-shear trapping\cite{NitscheHinch_1997}. The predicted  inhomogeneity is weak, however, and has not been observed in experiments.}; $\kappa$ is the swimmer aspect ratio and $Pe_r= U_m/(HD_r)$ is the rotary Peclet number measuring the relative importance of shear-induced rotation of $O(U_m/H)$, and relaxation due to rotary diffusion with diffusivity $D_r$, for active Brownian swimmers\,(ABP's). For large $Pe_r$ and slender prolate swimmers\,($\kappa \gg 1$), the trapping regimes are demarcated by $Pe_r \approx 0.07\kappa^3$\cite{vennamneni2020shear}; low-shear trapping, characterized by swimmers drifting towards the channel centerline, occurs for $Pe_r > 0.07\kappa^3$. 

Of particular significance to the active Taylor dispersion problem is the centerline collapse that occurs within the low-shear trapping regime for $\kappa \gtrsim 2$\cite{vennamneni2020shear}. For such swimmers, the shear-induced drift towards the centerline is strong enough for the swimmer concentration profile to approach a Dirac delta function in the limit $Pe_r \rightarrow \infty$\footnote{The multiple time scales framework used here assumes the swimmer mean free path to be negligibly small, leading to the swimmer concentration asymptoting to a Dirac delta function, for $Pe_r \rightarrow \infty$, in the centerline-collapse regime,. In practice, swimmers will localize in a near-center `Knudsen' layer of a finite thickness.}. This restricts the velocity variation sampled with increasing $Pe_r$, leading to an anomalously reduced dispersion. We map out classical and anomalous scaling regimes for $D_{ef\!f}$, for large $Pe_r$, as a function of $\kappa$, for prolate spheroidal swimmers\,($\kappa > 1$). Here, `classical' refers to cases where the $D_{ef\!f}$-scaling may be obtained as in the passive scenario, the only change being the replacement of $D$ by the $Pe_r$-dependent swimmer translational diffusivity; `anomalous' refers to cases where the centerline collapse leads instead to `tail events' controlling the reduced large-$Pe_r$-dispersion.

The probability density $\Omega(\mathbf{x},\mathbf{p},t)$ governing the evolution of a population of spheroidal ABP's satisfies the following kinetic equation\cite{SubKoch_2009,SS_2015,vennamneni2020shear,ConcBanding_2021}:
\begin{equation}
\hspace*{-0.2in}\frac{\partial \Omega}{\partial t}\!\!+\! \nabla_{\mathbf{x}}\!\cdot\![(\epsilon\mathbf{p}\!+\!Pe_r\mathbf{u})\Omega]\!+\!Pe_r \nabla_{p}\!\cdot\!(\dot{\mathbf{p}}\Omega)\!-\!\nabla^{2}_{p}\Omega\!=\!0, \label{swim:KineticEqn}
\end{equation}
where time\,($t$) and lengths\,($\mathbf{x}$) are scaled using $D_r^{-1}$ and $H$. The second term in (\ref{swim:KineticEqn}) denotes convection due to swimming\,($U_s\mathbf{p}$, $\mathbf{p}$ being the swimmer orientation and $U_s$ the swim speed), and the plane Poiseuille flow, $u_1(z)=(1-z^2)$, in non-dimensional form; $\epsilon = U_s/(HD_r)$, the ratio of the swimmer mean free path to channel half-width, may be regarded as a swimmer Knudsen number\cite{vennamneni2020shear,ConcBanding_2021}. The third and fourth terms denote shear-induced rotation and relaxation to isotropy due to rotary diffusion. The former is given by the Jeffery equation $\dot{\mathbf{p}}\!=\!B\!\left[ \mathbf{E}\!\cdot\! \mathbf{p}\!-\!\mathbf{p}(\mathbf{E}\!:\!\mathbf{p}\mathbf{p})\right]\!+\!\mathbf{W}\!\cdot\!\mathbf{p}$ with $B\!=\!(\kappa^{2}\!-\!1)/(\kappa^{2}\!+\!1)$\cite{Jeffery1922,HinchLeal1971}, $E_{ij}\!=\!-z(\delta_{i1}\delta_{j3}\!+\!\delta_{i3}\delta_{j1})$ and $W_{ij}\!=\!-z(\delta_{i1}\delta_{j3}\!-\!\delta_{i3}\delta_{j1})$ being the strain rate and vorticity tensors for plane Poiseuille flow.

For $\epsilon \ll 1$, time scales corresponding to orientation relaxation and drift/diffusion along the gradient and flow directions are well separated. The swimmer orientation distribution relaxes to an anisotropic quasi-steady state corresponding to the local shear rate in a time of $O(D_r^{-1})$, which governs subsequent spatial redistribution in the gradient direction on time scales of $O(H^2/D_t)$, or $O(D_r^{-1}\epsilon^{-2})$\,(using $D_t \sim U_s^2D_r^{-1}$ for the translational diffusivity\footnote{These are nominal estimates for $Pe_r \sim O(1)$; actual estimates involve additional $Pe_r$-dependent factors in both high and low-shear trapping regimes\cite{vennamneni2020shear}.}). On longer time scales of $O(L^2/D_{ef\!f})$,  $D_{ef\!f}$ being the shear-enhanced longitudinal diffusivity given by (\ref{Taylordiff}) below, the swimmer concentration settles to a steady function of $z$, and further evolution is via an $x-$dependent amplitude that determines the spread of the population along the flow direction. Here, $L$ is a characteristic longitudinal scale beyond which the swimmer population evolves diffusively, and may be inferred posteriori from the requirement $L^2/D_{ef\!f} \gg H^2/D_t$ imposed by the ordering of time scales above; for the classical case, $D_t = D$, $D_{ef\!f} \sim Pe^2D$, leading to $L \gg Pe\,H$ for $Pe \gg 1$.

The above aspects are captured by a multiple scales analysis where $\Omega(\mathbf{x},\mathbf{p},t)$ is expanded as:
\begin{align}
\hspace*{-0.2in}\Omega(\mathbf{x},\mathbf{p},t)\!\!=&K(x,t_2,t_3)\!F(z,t_2)G(\mathbf{p},t_1)  \nonumber \\
&+\!\epsilon \Omega^{(1)}(\mathbf{x},\mathbf{p},t_1,t_2,t_3) +O(\epsilon^2). \label{MS_expansion}
\end{align}
In (\ref{MS_expansion}), $t_1 = t$, $t_2 = \epsilon^2 t$ and $t_3 = \Lambda^2 \epsilon^2t$ denote progressively slower time scales in accordance with the aforementioned hierarchy of evolutions\cite{SubBrady_2004,vennamneni2020shear}. In contrast to (\ref{swim:KineticEqn}), $x$ and $z$ are now scaled with $H$ and $L$ respectively, and $\Lambda= H/L$ denotes the (small)\,ratio of the channel half-width to the appropriate longitudinal scale. Assuming spatial homogeneity along the flow direction, a version of (\ref{MS_expansion}) with only the $t_1$ and $t_2$-based evolutions leads to $F(z,t_2)$ satisfying a drift-diffusion equation\cite{vennamneni2020shear,SM}:
\begin{equation}
\frac{\partial F}{\partial t_2} + \frac{\partial }{\partial z}(V_z F)= \frac{\partial}{\partial z}\left( D_{zz} \frac{\partial F}{\partial z}\right), \label{Gradientdiff}
\end{equation}
with $V_z(z;Pe_r,\kappa)$ and $D_{zz}(z;Pe_r,\kappa)$ being the gradient component of the drift and translational diffusivity, respectively, which determine the swimmer distribution along the gradient direction. The normalized steady state concentration profile is $F_s(z;Pe_r,\kappa)  \propto \exp[\textstyle\int_{-1}^z dz'\frac{V_z(z')}{D_{zz}(z')}]$. For $\kappa$ fixed, a change in sign of $V_z$ leads to a transition from high to low-shear trapping with increasing $Pe_r$\cite{vennamneni2020shear}.

For a population finite in extent along the flow direction, relevant to the dispersion problem, an $x$ and a slower time\,($t_3$) dependence are included via the amplitude function $K$ in (\ref{MS_expansion}). It may be shown that $F$ still satisfies (\ref{Gradientdiff}), while $K(x,t_2,t_3)\equiv K(\bar{x},t_2)$ satisfies a diffusion equation, $\frac{\partial K}{\partial t_3} = D_{ef\!f} \frac{\partial^2 K}{\partial \bar{x}^2}$ in a reference frame moving with the mean swimmer speed defined by $\bar{u}_1(Pe_r)\!=\!\textstyle\int_{-1}^1 u_1 F_sdz$; $\bar{x}\!=\!x\!-\!\bar{u}_1 t_2$. The effective diffusivity in this equation is given by\cite{SM}:
\begin{equation}
\hspace*{-0.2in}\frac{D_{ef\!f}}{D_t}\!\!=\!\frac{2Pe_r^2}{\epsilon^4}\!\!\! \displaystyle\int_{0}^1\!\! \frac{dz}{F_s(z)D_{zz}(z)}\!\!\left[ \displaystyle\int_{0}^z \!\!\!dz' u'_1(z')F_s(z')\!\right]^2\!\!\!, \label{Taylordiff}
\end{equation}
for large $Pe_r$, with $u_1'\!=\!u_1 -\bar{u}_1$ denoting the longitudinal velocity fluctuation; note that $Pe = Pe_r/\epsilon^2$, whence the prefactor in (\ref{Taylordiff}) is $2Pe^2$. Using $D_{zz} = 1, F_s =\frac{1}{2}$ leads to the classical $O(Pe^2)$ contribution\cite{Wooding_1960}.

\begin{figure}
\centering
\includegraphics[height=55mm,width=1.\columnwidth]{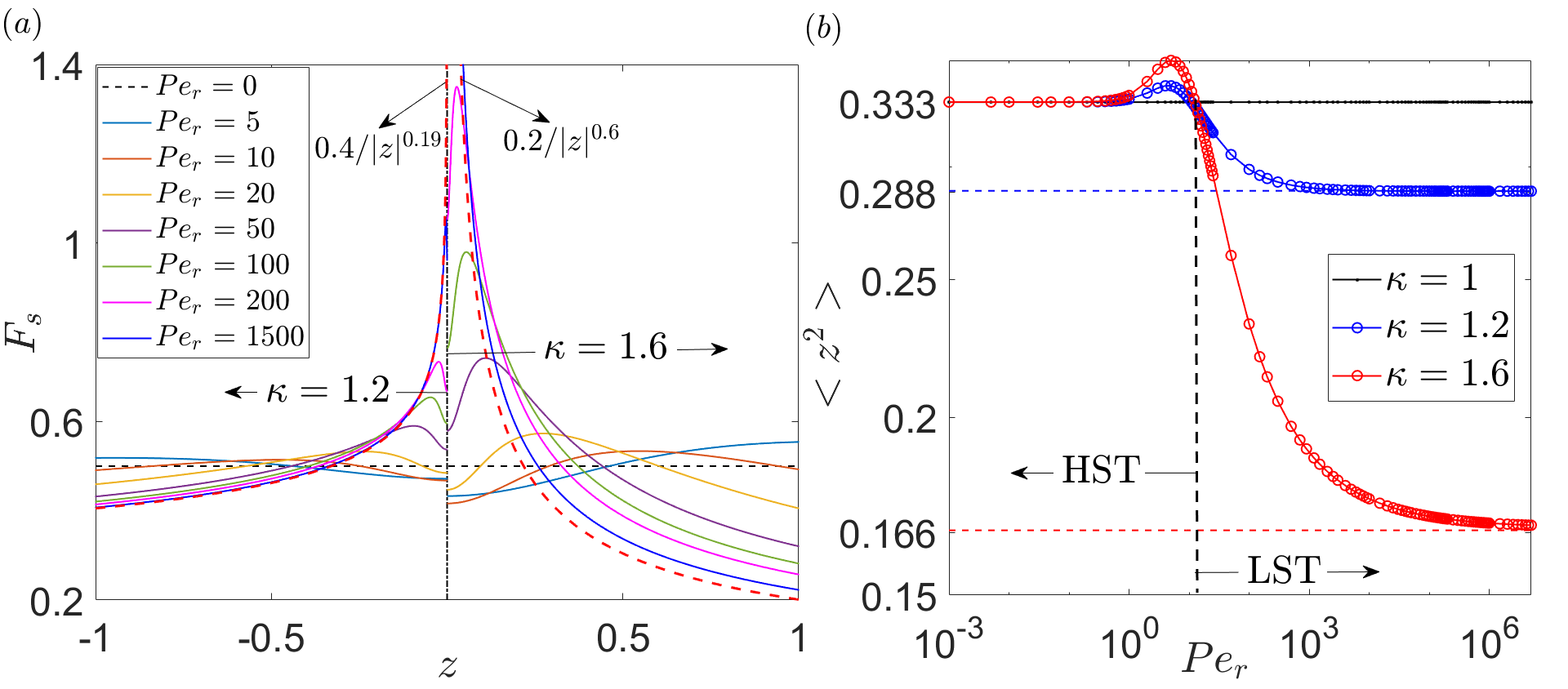}
\includegraphics[height=55mm,width=1.\columnwidth]{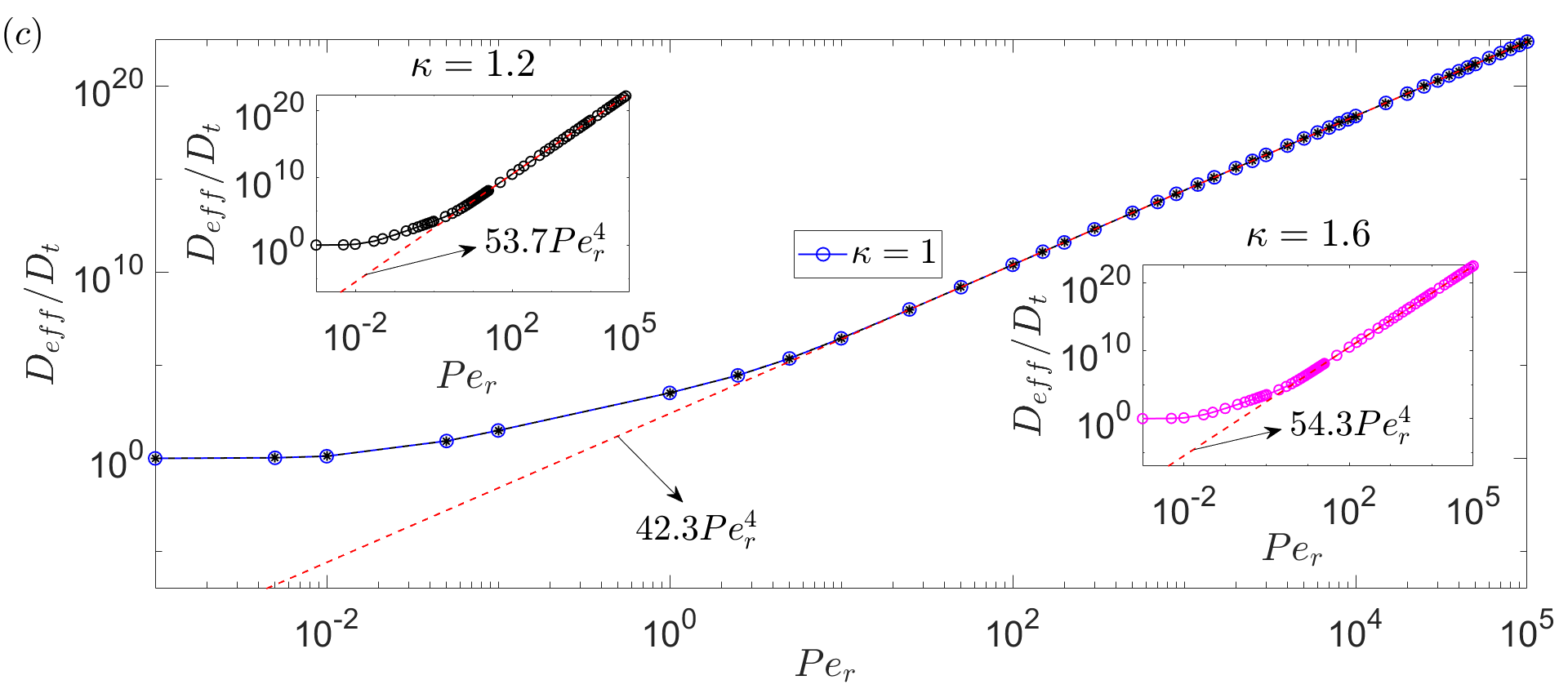}
\caption{(a) Swimmer concentration profiles for different $Pe_r$:$\kappa\!=\!1.2$(left),$1.6$(right)- the infinite-$Pe_r$ profiles appear as dashed red curves; (b)$\langle z^2 \rangle\!=\! \textstyle\int_{-1}^1\!\!z^2 F_s(z) dz$ as a function of $Pe_r$ for $\kappa = 1.2$ and $1.6$--LST$\equiv$low-shear trapping, HST$\equiv$high-shear trapping; (c) $D_{ef\!f}/D_t$ v\!/\!s $Pe_r$ for $\kappa\!=\!1$\,($\kappa =1.2$, $1.6$ appear as insets); red dashed lines are predictions from (\ref{axialdiffu:kappaless2})\,($C=0,C_1\!\approx\!8/3$ for $\kappa\!=\!1$; $C\! \approx\!-0.19,C_1\!\approx\!2.12$ for $\kappa\!=\!1.2$; $C\!\approx\!-0.6,C_1\!\approx\!1.73$ for $\kappa\!=\!1.6$)}
\label{fig:td2}
\end{figure}
Fig\,\ref{fig:td2}a illustrates the transition of the swimmer concentration profiles at a given $x$\,($t_2 \gg 1$), from a high-shear to a low-shear trapping regime with increasing $Pe_r$, for a pair of $\kappa$'s\,($1.2$ and $1.6$) below the centerline-collapse threshold\cite{vennamneni2020shear}. Both sets of profiles approach $Pe_r$-independent limiting forms for $Pe_r \rightarrow \infty$, the profile variances, $\langle z^2\rangle\!=\!\textstyle\int_{-1}^1 z^2 F_s(z;Pe_r,\kappa) dz$, asymptoting to finite plateaus for large $Pe_r$\,(Fig\,\ref{fig:td2}b). The variance is a non-monotonic function of $Pe_r$, reflecting an initial increase due to high-shear trapping, followed by the eventual decrease due to swimmers drifting towards the centerline; both the amplitude of the initial high-shear trapping, and that of the subsequent decrease, are larger for $\kappa = 1.6$. $D_{zz}$ and $V_z$ scale as $O(Pe_r^{-2})$ for sufficiently large $Pe_r$\cite{vennamneni2020shear}, and $F_s$, which depends only on their ratio, approaches $F_s^\infty(z;\kappa) = [1\!+\!C(\kappa)]|z|^{C(\kappa)}/2$\cite{vennamneni2020shear}. $C = 0$ for $\kappa = 1$, so spherical swimmers are uniformly distributed, with $Pe_r$ only determining the rate of approach to the uniform state\footnote{The uniform state is the result of neglecting direct interactions of the swimmers, hydrodynamic or otherwise, with boundaries; such interactions lead to localized wall layers which play a significant role for narrow channels corresponding to finite $\epsilon$\cite{Gompper1,Gompper2,Underhill2015}.}; the $F_s^\infty$'s for $\kappa = 1.2$ and $1.6$ are indicated in Fig\,\ref{fig:td2}a. Using $F^\infty_s$ above, and the large-$Pe_r$ asymptote, $D_{zz} \approx C_1/(2zPe_r)^2$\cite{vennamneni2020shear}, in (\ref{Taylordiff}), one obtains:
\begin{equation}
\hspace*{-0.25in}\lim_{Pe_r \rightarrow \infty}\!\!\! \frac{D_{ef\!f}}{D_t}\!\!=\!\epsilon^{-4}\!Pe_r^4\!\frac{32(1\!+\!C)}{C_1\!(C\!+\!3)^2(315\!+\!143C\!+\!21C^2\!+\!C^3\!)}, \label{axialdiffu:kappaless2}
\end{equation}
with $D_t = U_s^2/6D_r$ sand $C> -1$ for $\kappa \lesssim 2$.

Fig \ref{fig:td2}c plots $D_{ef\!f}/D_t$, approximated as the sum of its zero-$Pe_r$\,(which equals unity) and large-$Pe_r$\,(given by (\ref{Taylordiff})) limiting forms, for $\kappa = 1, 1.2$ and $1.6$, the latter two as insets. $D_{ef\!f}/D_t$ for $\kappa =1$ may be obtained in closed form for arbitrary $Pe_r$\cite{SM}. In all cases, $D_{ef\!f}/D_t$ starts from an initial plateau and eventually increases as $Pe_r^4$, the latter scaling
readily explained using the classical argument\,(introductory paragraph) with the $Pe_r^{-2}$ scaling of $D_{zz}$. While this scaling has been identified earlier for spherical swimmers\cite{GuastoPNAS2019}, Fig \ref{fig:td2}c shows that it holds for swimmers with $1\!\leq\!\kappa\!\lesssim\!2$. Predictions based on (\ref{axialdiffu:kappaless2}), which give the numerical prefactor in the $Pe_r^4$-asymptote, agree with those obtained from evaluating (\ref{Taylordiff}) numerically. This prefactor is a non-monotonic function of $\kappa$: it equals $42.3$ for $\kappa = 1$, increases to about $59$ for $\kappa \approx 1.4$, before decreasing to zero for $\kappa \approx 2$\,(see below). Swimmers with $\kappa \approx 1.4$ are therefore the fastest dispersing prolate swimmers.

\begin{figure}
\centering
\includegraphics[height=55mm,width=1.\columnwidth]{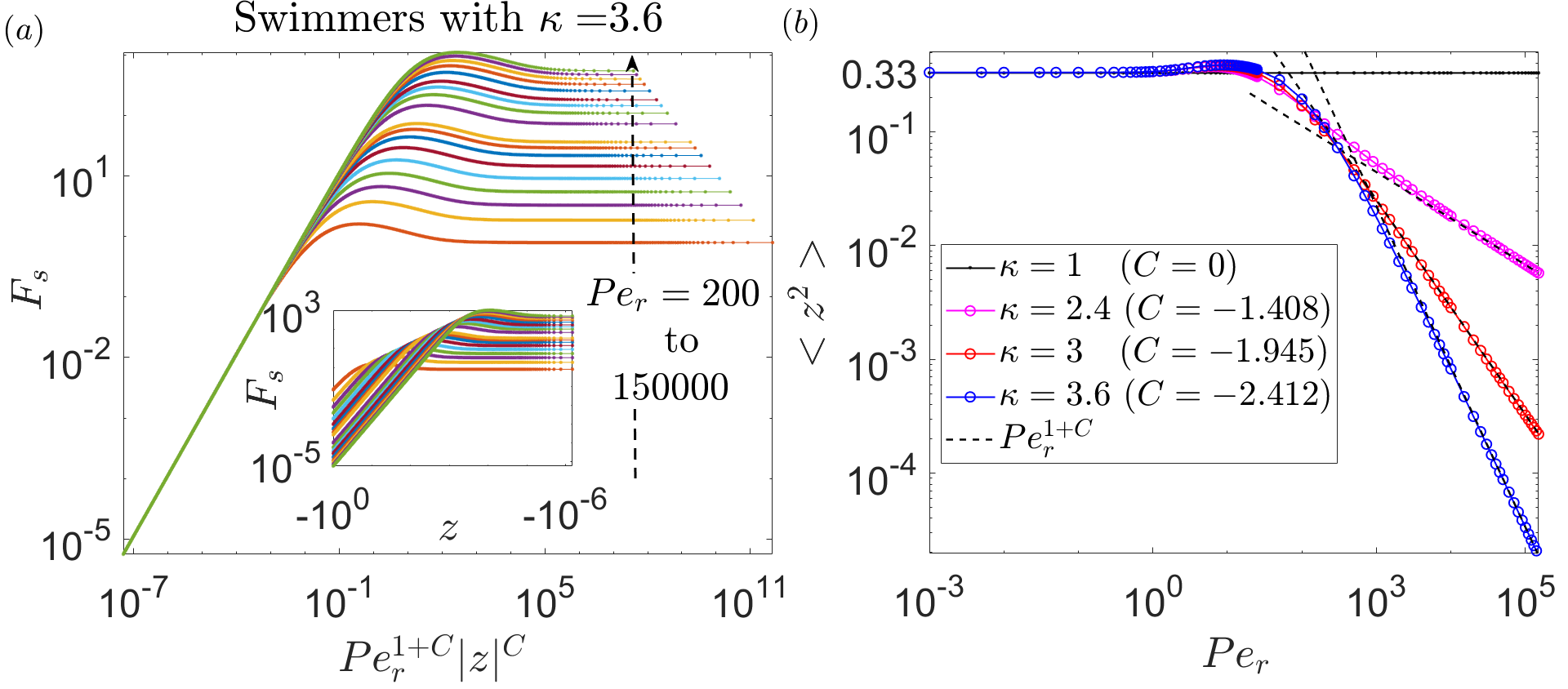}
\includegraphics[height=55mm,width=1.\columnwidth]{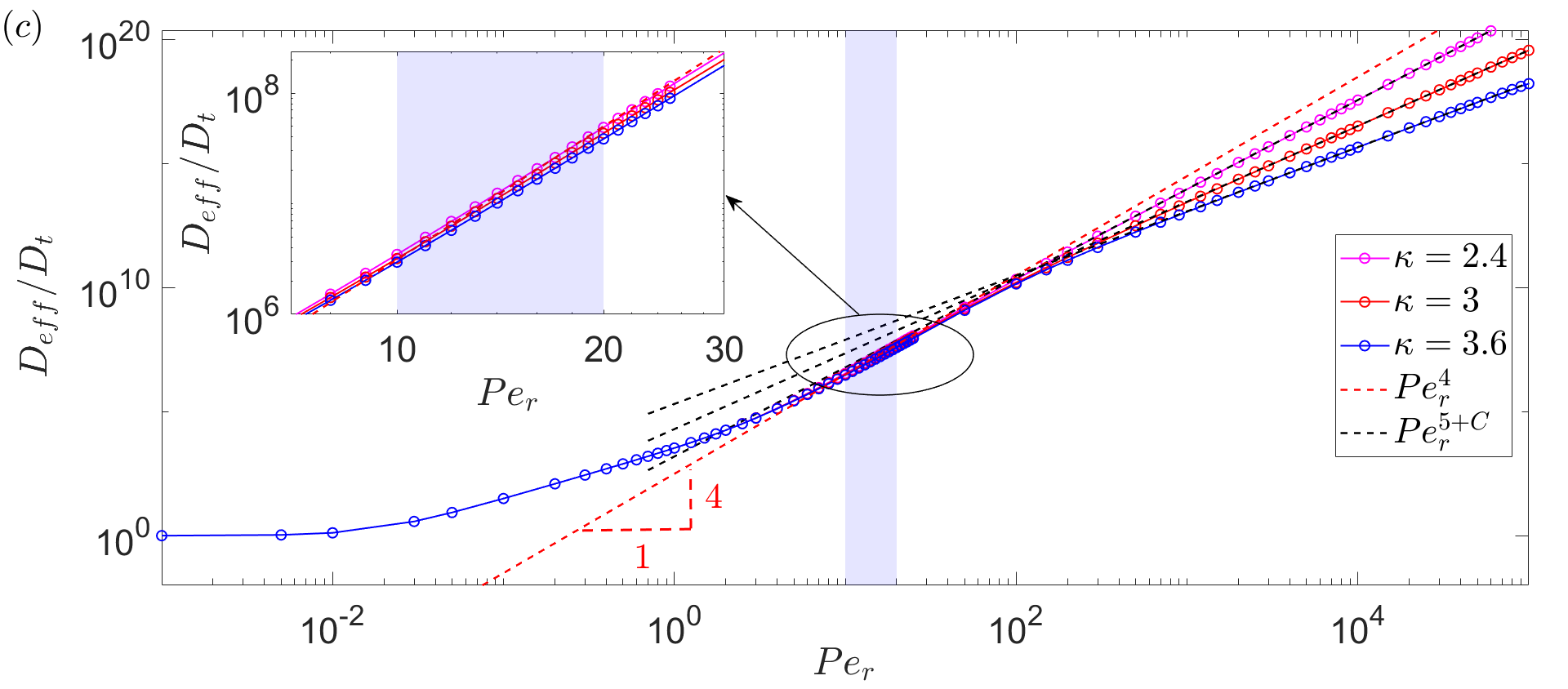}
\caption{(a) Collapsed tails of scaled concentration profiles for $\kappa =3.6$\,(inset shows unscaled profiles); (b) $\langle z^2 \rangle$, for $\kappa = 2.4$, $3$ and $3.6$, decaying as $Pe_r^{1+C}$ for $Pe_r \rightarrow \infty$; (c) $D_{ef\!f}/D_t$ v\!/\!s $Pe_r$ for the same $\kappa$'s; the anomalous asymptotes\,($\propto Pe_r^{5+C}$) appear as dashed black lines, while the classical $Pe_r^4$-scaling for intermediate $Pe_r$ appears as a dashed red line; the vertical band marks the initial departure from the classical regime.}
\label{fig:td3}
\end{figure}

$C(\kappa)$ dips below $-1$ for $\kappa \gtrsim 2$, and $F_s^\infty$ becomes non-normalizable as a result. Swimmers with $\kappa \gtrsim 2$ collapse onto the centerline\cite{vennamneni2020shear}, with $F_s(z;Pe_r,\kappa)$ approaching $\delta(z)$ for $Pe_r \rightarrow \infty$. As shown in Fig \ref{fig:td3}a, for $\kappa = 3.6$, $F_s\!\propto\!Pe_r^{1+C}|z|^C$ outside a shrinking $O(Pe_r^{-1})$ central core where it saturates at an $O(Pe_r)$ value. The profile variances in Fig \ref{fig:td3}b, following an initial increase, approach zero as $Pe_r^{1+C}$ for $Pe_r \rightarrow \infty$. The implications for dispersion are seen from (\ref{axialdiffu:kappaless2}) where the coefficient of $Pe_r^4$ approaches zero for $\kappa \rightarrow 2^-\,(C\rightarrow -1^+)$, suggesting that $D_{ef\!f}$ must eventually increase as a smaller power of $Pe_r$ for $\kappa \gtrsim 2$. This is confirmed in Fig \ref{fig:td3}c for $\kappa$'s in the interval $2\!-\!4$. $D_{ef\!f}/D_t$ now shows a first transition to a $Pe_r^4$-scaling regime, followed by a second transition to an anomalous regime characterized by an exponent less than $4$, and that decreases with increasing $\kappa$. For the chosen $\kappa$'s, the initial departure from the classical scaling happens over the interval $Pe_r \sim 10-20$\,(marked by a vertical band in Fig.\ref{fig:td3}c; also see inset).

The anomalously reduced dispersion in Fig \ref{fig:td3}c is readily rationalized based on the vanishingly small velocity variations sampled by a swimmer, localized in a shrinking neighborhood of the centerline, for $Pe_r \rightarrow \infty$. An argument for the scaling exponent along classical lines, with a channel width of $O(HPe_r^{-1})$ corresponding to the core region, does not work. The dispersion is now controlled by the small fraction of swimmers outside the core that sample the largest velocity variations. To see this, we estimate contributions to the integral in (\ref{Taylordiff}) from core\,($|z| \sim O(Pe_r^{-1})$) and tail\,($z \sim O(1)$) regions. The majority of the swimmers are in the core, and the average swimmer speed is therefore unity at leading order. Accounting for the $O(Pe_r^{1+C})$ fraction of swimmers outside, with $O(1)$ velocity deficits, leads to the more accurate estimate $\bar{u}_1\!\sim\!1-O(Pe_r^{1+C})$. One therefore has $z' \sim O(Pe_r^{-1}), F_s \sim O(Pe_r), u'_1 \sim O(Pe_r^{1+C})$ in the core, leading to the inner integral $\textstyle\int_0^{O(Pe_r^{-1})}\!u'_1F_s dz'$ in (\ref{Taylordiff}) being $O(Pe_r^{1+C})$; using $D_{zz} \sim O(1)$ owing to the vanishingly small shear in the core, the core contribution, $\textstyle\int_0^{O(Pe_r^{-1})}\!\!\frac{dz}{F_sD_{zz}}.O[Pe_r^{2(1+C)}]$, is $O(Pe_r^{2C})$. In the tail, $z' \sim O(1), F_s \sim O(Pe_r^{1+C}), u'_1 \sim O(1)$. The inner integral is again $O(Pe_r^{1+C})$, as must be, since the net swimmer flux is zero in the chosen reference frame, implying equal and opposite core and tail region contributions. Using $D_{zz} \sim O(Pe_r^{-2}), z \sim O(1)$, one obtains the dominant tail contribution as $O(Pe_r^{3+C})$. Accounting for the prefactor in (\ref{Taylordiff}), $D_{ef\!f}/D_t \propto Pe_r^{5+C(\kappa)}\!\epsilon^{-4}$ for $\kappa\!\gtrsim\!2$, which agrees well with the numerical curves\,(Fig \ref{fig:td3}c). For $\kappa \approx 2$, one expects  $D_{ef\!f}/D_t \propto Pe_r^4\epsilon^{-4}(\ln Pe_r)^{-1}$, although the logarithmic scaling is difficult to verify, given that the threshold $\kappa$ is only known approximately.

The plots of $D_{ef\!f}/D_t$ for larger aspect ratios upto $\kappa = 12$, in Fig \ref{fig:td4}, indicate postponement of the onset of anomalous scaling to larger $Pe_r$ as $\kappa$ increases\,(the black dots in the upper inset correspond to transition $Pe_r$'s increasing from $634$ for $\kappa = 5$ to $3142$ for $\kappa = 12$). Since onset correlates to low-shear trapping and the associated centerline collapse, one expects $Pe_r$ at onset to eventually scale as $\kappa^3$ for $\kappa \gg 1$\cite{vennamneni2020shear}, although one needs to go to larger $\kappa$ to observe this scaling. A second feature is the emergence, for the largest $\kappa$'s, of an intermediate asymptotic regime with $D_{ef\!f}/D_t \propto Pe_r^{\frac{10}{3}}$, prior to the anomalous one. This regime pertains to slender swimmers that exhibit high-shear trapping for $1\ll  Pe_r \ll \kappa^3$ with $D_{zz}, V_z \sim O(Pe_r^{-\frac{4}{3}})$\cite{vennamneni2020shear}- this $D_{zz}$-scaling, used in the classical argument, yields the exponent\,($10/3$) above. Thus, $\kappa \rightarrow \infty$ is a singular limit when the anomalous regime recedes to infinity, and $D_{ef\!f}/D_t$ shows a single transition from the initial plateau to the $Pe_r^{\frac{10}{3}}$-scaling regime\,(dash-dotted line in Fig {\ref{fig:td4}). Clearly, the infinite aspect ratio approximation, used in earlier efforts\cite{Bearon_2015,Saintillan_2015}, does not capture the large-$Pe_r$ dispersion behavior for large but finite $\kappa$. 
\begin{figure}
\centering
\includegraphics[height=75mm,width=1\columnwidth]{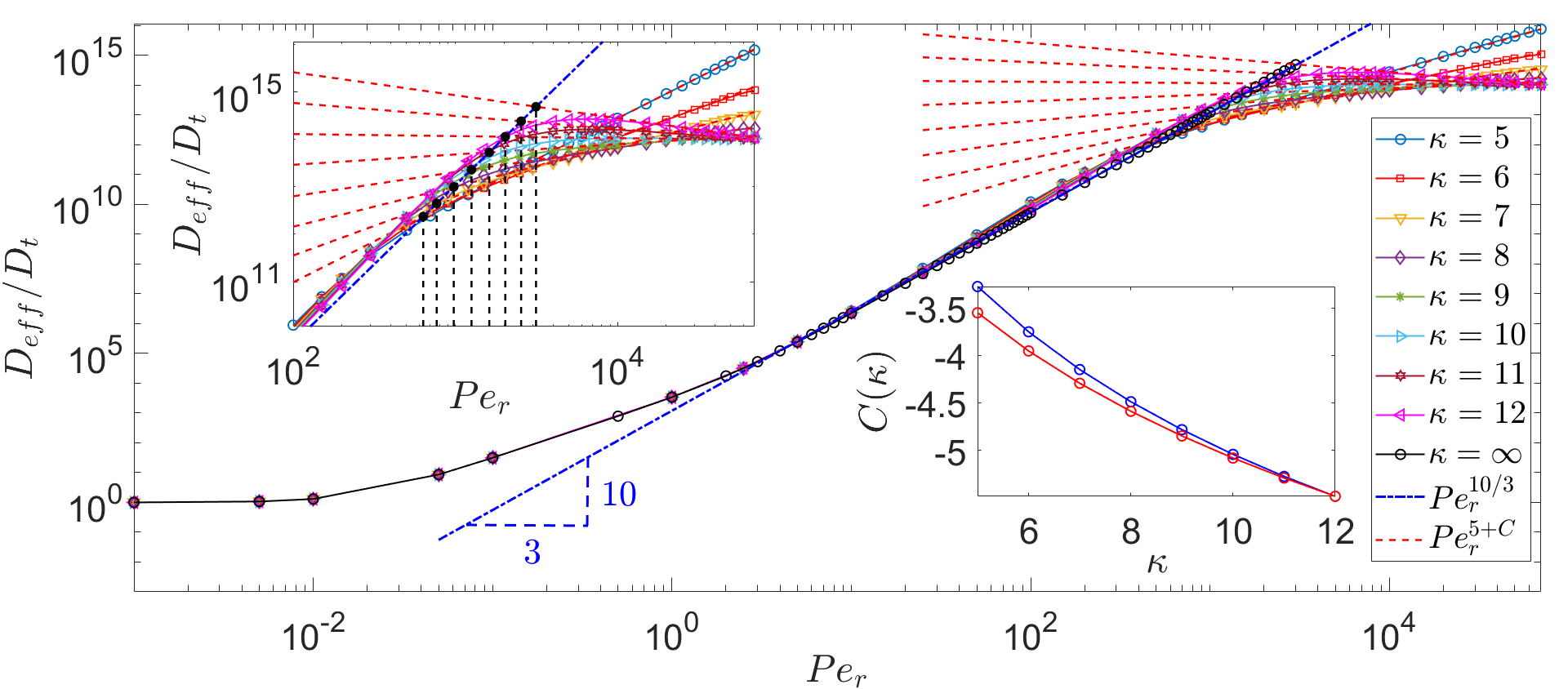}
\caption{$D_{ef\!f}/D_t$\,v\!/\!s\,$Pe_r$  for $\kappa\!=\!5\!-\!12$; anomalous asymptotes appear as dashed red lines, and the intermediate-regime $Pe_r^{\frac{10}{3}}$-asymptote as a dash-dotted blue line. Black dots and vertical dashed lines in the upper inset, corresponding to the intersection of these two asymptotes, mark the transition to anomalous scaling; lower inset characterizes the large-$\kappa$ scaling of $C(\kappa)$.}
\label{fig:td4}
\end{figure}

There are a couple of noteworthy points in Fig \ref{fig:td4}. First, the decrease in $C(\kappa)$ causes the anomalous exponent to dip below $2$ for $\kappa \gtrsim 4.5$, implying that the centerline-collapse of such swimmers leads to their dispersion, for large $Pe_r$, becoming slower than passive particles with the same aspect ratio. For these swimmers, a source of translational diffusion unrelated to swimming activity\,(corresponding to an additional term $\hat{D}_t \nabla^2_{\mathbf{x}} \Omega$ in (\ref{swim:KineticEqn}), with $\hat{D}_t$ independent of $D_r$) implies a third transition from the anomalous $Pe_r^{5+C}$\!\!-scaling to a classical $Pe_r^2$ one for sufficiently large $Pe_r$; although, the transition $Pe_r$ is likely unrealistically large if $\hat{D}_t$ has a thermal origin. A second more remarkable feature is that $C \approx -5$ for $\kappa \approx 10$, decreasing further for larger $\kappa$, and approaching an $O(\ln \kappa)$ scaling for $\kappa \rightarrow \infty$\,(lower inset in Fig \ref{fig:td4}). This implies a near-zero anomalous exponent for $\kappa \approx 10$, and a negative one for larger $\kappa$, implying a decrease in $D_{ef\!f}/D_t$ with increasing $Pe_r$ for $\kappa \gtrsim 10$. Thus, for $\kappa \gtrsim 10$, $D_{ef\!f}/D_t$ starts off at the low-$Pe_r$ plateau and first increases as $Pe_r^{\frac{10}{3}}$, before decreasing as $Pe^{5+C}$ for larger $Pe_r$. $D_{ef\!f}$ attains a maximum at the transition between the latter two regimes, so the large-$Pe_r$-dispersion of  swimmers with $\kappa \gtrsim 10$ is limited by this maximum value. Note that, for sufficiently slender swimmers\,($\kappa \gg 1$), the transition $Pe_r$ is $O(\kappa^3)$, so the aforementioned maximum is $O(\kappa^{10}D_t)$; although, achieving this scaling requires going to $\kappa$ larger than $12$.

By examining prolate spheroidal swimmers in plane Poiseuille flow, we have characterized the influence of activity and swimmer anisotropy on longitudinal dispersion. Activity is shown to both enhance, and importantly, reduce dispersion, the reduction driven by the singular collapse of swimmers towards the centerline for large $Pe_r$. The centerline collapse manifests as an anomalous dispersion exponent, determined by large velocity fluctuations experienced by swimmers during rare excursions outside an $O(Pe_r^{-1})$ central core. For $\kappa \sim O(1)$, the departure from classical scaling already sets in at $Pe_r \approx 10$, being very relevant to experiments. While the importance of tail-events, not accounted for by $D_{ef\!f}$, is known for the classical Taylor dispersion problem\cite{Haynes_2014}, and for other problems in statistical physics including turbulence\cite{KRS_1997}, their relevance in the determining $D_{ef\!f}$ itself for active swimmers is interesting. Several recent efforts have focused on the wallward migration of swimmers, and the associated upstream swimming\cite{Hill_2007,Costanzo_2012,Kaya_2012,Lindner_2015,Saintillan_2015}. The work here shows that there exist independent and non-trivial bulk mechanisms that drive migration away from boundaries, and can potentially affect both incipient bio-film formation and rate of spread of microbial populations.

The dispersion behavior also depends on the overall swimmer geometry, for instance, on whether swimmers have a rod-like or disk-like morphology. While discoid swimmers\,(modeled as oblate spheroids with $\kappa < 1$) exhibit a reverse transition from low to high-shear trapping with increasing $Pe_r$\cite{ProlateOblateman}, there isn't an analogous wall-collapse for $Pe_r \rightarrow \infty$. For flat disk swimmers\,($\kappa \ll 1$), $D_{ef\!f}$ transitions from a zero-$Pe_r$ plateau, via an intermediate asymptotic regime with $D_{ef\!f}/D_t\!\propto\!Pe_r^{\frac{8}{3}}$, to an eventual $Pe_r^4$-scaling regime, all of these exponents obtainable using classical arguments\cite{ProlateOblateman}. 

The dispersion behavior found here is relevant to general unidirectional shearing flows. Although the exponent values will differ, the onset of anomalous scaling and the eventual geometrically limited dispersion behavior, for large $\kappa$, will also occur in pipe Poiseuille flow; the centerline-collapse threshold remains $\kappa \approx 2$\cite{vennamneni2020shear}. To the extent that the Poiseuille profile is a local approximation for a velocity extremum, one expects the dispersion of rod-like swimmers in unidirectional shearing flows to be limited by collapse onto velocity extrema. In this regard, it is of particular interest to investigate the effect, on centerline collapse and the implication for dispersion, of the onset of bacterial turbulence\cite{SS2007,SS2008,SubKoch_2009,KochSub_2011,Deepak_2015,Martinez_2020}. The dispersion of a passive solute in turbulent flow was examined by Taylor himself\cite{Taylor_1954}, although the implications for active swimmers appear much more intriguing.
  
\bibliography{References}

\begin{thebibliography}{43}%
\makeatletter
\providecommand \@ifxundefined [1]{%
 \@ifx{#1\undefined}
}%
\providecommand \@ifnum [1]{%
 \ifnum #1\expandafter \@firstoftwo
 \else \expandafter \@secondoftwo
 \fi
}%
\providecommand \@ifx [1]{%
 \ifx #1\expandafter \@firstoftwo
 \else \expandafter \@secondoftwo
 \fi
}%
\providecommand \natexlab [1]{#1}%
\providecommand \enquote  [1]{``#1''}%
\providecommand \bibnamefont  [1]{#1}%
\providecommand \bibfnamefont [1]{#1}%
\providecommand \citenamefont [1]{#1}%
\providecommand \href@noop [0]{\@secondoftwo}%
\providecommand \href [0]{\begingroup \@sanitize@url \@href}%
\providecommand \@href[1]{\@@startlink{#1}\@@href}%
\providecommand \@@href[1]{\endgroup#1\@@endlink}%
\providecommand \@sanitize@url [0]{\catcode `\\12\catcode `\$12\catcode
  `\&12\catcode `\#12\catcode `\^12\catcode `\_12\catcode `\%12\relax}%
\providecommand \@@startlink[1]{}%
\providecommand \@@endlink[0]{}%
\providecommand \url  [0]{\begingroup\@sanitize@url \@url }%
\providecommand \@url [1]{\endgroup\@href {#1}{\urlprefix }}%
\providecommand \urlprefix  [0]{URL }%
\providecommand \Eprint [0]{\href }%
\providecommand \doibase [0]{http://dx.doi.org/}%
\providecommand \selectlanguage [0]{\@gobble}%
\providecommand \bibinfo  [0]{\@secondoftwo}%
\providecommand \bibfield  [0]{\@secondoftwo}%
\providecommand \translation [1]{[#1]}%
\providecommand \BibitemOpen [0]{}%
\providecommand \bibitemStop [0]{}%
\providecommand \bibitemNoStop [0]{.\EOS\space}%
\providecommand \EOS [0]{\spacefactor3000\relax}%
\providecommand \BibitemShut  [1]{\csname bibitem#1\endcsname}%
\let\auto@bib@innerbib\@empty
\bibitem [{\citenamefont {Taylor}(1953)}]{Taylor_1953}%
  \BibitemOpen
  \bibfield  {author} {\bibinfo {author} {\bibfnamefont {G.}~\bibnamefont
  {Taylor}},\ }\href@noop {} {\bibfield  {journal} {\bibinfo  {journal} {Prof.
  R. Soc. Lond.}\ }\textbf {\bibinfo {volume} {A219}},\ \bibinfo {pages} {186}
  (\bibinfo {year} {1953})}\BibitemShut {NoStop}%
\bibitem [{\citenamefont {Aris}(1956)}]{Aris_1956}%
  \BibitemOpen
  \bibfield  {author} {\bibinfo {author} {\bibfnamefont {R.}~\bibnamefont
  {Aris}},\ }\href@noop {} {\bibfield  {journal} {\bibinfo  {journal} {Prof. R.
  Soc. Lond.}\ }\textbf {\bibinfo {volume} {A235}},\ \bibinfo {pages} {67}
  (\bibinfo {year} {1956})}\BibitemShut {NoStop}%
\bibitem [{\citenamefont {Wooding}(1960)}]{Wooding_1960}%
  \BibitemOpen
  \bibfield  {author} {\bibinfo {author} {\bibfnamefont {R.}~\bibnamefont
  {Wooding}},\ }\href@noop {} {\bibfield  {journal} {\bibinfo  {journal}
  {Journal of Fluid Mechanics}\ }\textbf {\bibinfo {volume} {7}},\ \bibinfo
  {pages} {501} (\bibinfo {year} {1960})}\BibitemShut {NoStop}%
\bibitem [{\citenamefont {Dorfman}\ and\ \citenamefont
  {Brenner}(2001)}]{Dorfman_2001}%
  \BibitemOpen
  \bibfield  {author} {\bibinfo {author} {\bibfnamefont {K.}~\bibnamefont
  {Dorfman}}\ and\ \bibinfo {author} {\bibfnamefont {H.}~\bibnamefont
  {Brenner}},\ }\href@noop {} {\bibfield  {journal} {\bibinfo  {journal} {J.
  Applied Phys.}\ }\textbf {\bibinfo {volume} {90}},\ \bibinfo {pages} {6553}
  (\bibinfo {year} {2001})}\BibitemShut {NoStop}%
\bibitem [{Note1()}]{Note1}%
  \BibitemOpen
  \bibinfo {note} {Unlike a linear flow\cite {Foister_1980,SubBrady_2004}, the
  mean position of the Brownian particle also evolves with time in a reference
  frame moving with $U_m$, with analogous arguments giving $<x> \sim -\protect
  \frac {U_m D}{H^2} t^2$ for short times, and $<x> \sim -U_m t$ for times
  longer than $O(H^2/D)$, reflecting the fact that the solute distribution
  moves with the mean speed rather than the centerline maximum. The variance,
  $<\protect \tmspace -\thinmuskip {.1667em}\protect \tmspace -\thinmuskip
  {.1667em}x^2\protect \tmspace -\thinmuskip {.1667em}\protect \tmspace
  -\thinmuskip {.1667em}>-<\protect \tmspace -\thinmuskip {.1667em}\protect
  \tmspace -\thinmuskip {.1667em}x\protect \tmspace -\thinmuskip
  {.1667em}\protect \tmspace -\thinmuskip {.1667em}>^2$, scales in the same
  manner as $<\protect \tmspace -\thinmuskip {.1667em}\protect \tmspace
  -\thinmuskip {.1667em}x^2\protect \tmspace -\thinmuskip {.1667em}\protect
  \tmspace -\thinmuskip {.1667em}>$.}\BibitemShut {Stop}%
\bibitem [{\citenamefont {Zottl}\ and\ \citenamefont
  {Stark}(2013)}]{Stark2013}%
  \BibitemOpen
  \bibfield  {author} {\bibinfo {author} {\bibfnamefont {A.}~\bibnamefont
  {Zottl}}\ and\ \bibinfo {author} {\bibfnamefont {H.}~\bibnamefont {Stark}},\
  }\href@noop {} {\bibfield  {journal} {\bibinfo  {journal} {Eur. Phys. J. E}\
  }\textbf {\bibinfo {volume} {36}},\ \bibinfo {pages} {4} (\bibinfo {year}
  {2013})}\BibitemShut {NoStop}%
\bibitem [{\citenamefont {Bearon}\ and\ \citenamefont
  {Hazel}(2015)}]{Bearon_2015}%
  \BibitemOpen
  \bibfield  {author} {\bibinfo {author} {\bibfnamefont {R.}~\bibnamefont
  {Bearon}}\ and\ \bibinfo {author} {\bibfnamefont {A.}~\bibnamefont {Hazel}},\
  }\href@noop {} {\bibfield  {journal} {\bibinfo  {journal} {J. Fluid Mech.}\
  }\textbf {\bibinfo {volume} {771}},\ \bibinfo {pages} {R3} (\bibinfo {year}
  {2015})}\BibitemShut {NoStop}%
\bibitem [{\citenamefont {Ezhilan}\ and\ \citenamefont
  {Saintillan}(2015)}]{Saintillan_2015}%
  \BibitemOpen
  \bibfield  {author} {\bibinfo {author} {\bibfnamefont {B.}~\bibnamefont
  {Ezhilan}}\ and\ \bibinfo {author} {\bibfnamefont {D.}~\bibnamefont
  {Saintillan}},\ }\href@noop {} {\bibfield  {journal} {\bibinfo  {journal} {J.
  Fluid Mech.}\ }\textbf {\bibinfo {volume} {777}},\ \bibinfo {pages} {482}
  (\bibinfo {year} {2015})}\BibitemShut {NoStop}%
\bibitem [{\citenamefont {Rusconi}\ \emph {et~al.}(2014)\citenamefont
  {Rusconi}, \citenamefont {Guasto},\ and\ \citenamefont
  {Stocker}}]{Rusconi_2014}%
  \BibitemOpen
  \bibfield  {author} {\bibinfo {author} {\bibfnamefont {R.}~\bibnamefont
  {Rusconi}}, \bibinfo {author} {\bibfnamefont {J.}~\bibnamefont {Guasto}}, \
  and\ \bibinfo {author} {\bibfnamefont {R.}~\bibnamefont {Stocker}},\
  }\href@noop {} {\bibfield  {journal} {\bibinfo  {journal} {Nat. Phys.}\
  }\textbf {\bibinfo {volume} {10}},\ \bibinfo {pages} {212} (\bibinfo {year}
  {2014})}\BibitemShut {NoStop}%
\bibitem [{\citenamefont {Barry}\ \emph {et~al.}(2015)\citenamefont {Barry},
  \citenamefont {Rusconi}, \citenamefont {Guasto},\ and\ \citenamefont
  {Stocker}}]{Barry_2015}%
  \BibitemOpen
  \bibfield  {author} {\bibinfo {author} {\bibfnamefont {M.}~\bibnamefont
  {Barry}}, \bibinfo {author} {\bibfnamefont {R.}~\bibnamefont {Rusconi}},
  \bibinfo {author} {\bibfnamefont {J.}~\bibnamefont {Guasto}}, \ and\ \bibinfo
  {author} {\bibfnamefont {R.}~\bibnamefont {Stocker}},\ }\href@noop {}
  {\bibfield  {journal} {\bibinfo  {journal} {J. R. Soc Interface}\ }\textbf
  {\bibinfo {volume} {12}},\ \bibinfo {pages} {20150791} (\bibinfo {year}
  {2015})}\BibitemShut {NoStop}%
\bibitem [{\citenamefont {Vennamneni}\ \emph
  {et~al.}(2020{\natexlab{a}})\citenamefont {Vennamneni}, \citenamefont
  {Nambiar},\ and\ \citenamefont {Subramanian}}]{vennamneni2020shear}%
  \BibitemOpen
  \bibfield  {author} {\bibinfo {author} {\bibfnamefont {L.}~\bibnamefont
  {Vennamneni}}, \bibinfo {author} {\bibfnamefont {S.}~\bibnamefont {Nambiar}},
  \ and\ \bibinfo {author} {\bibfnamefont {G.}~\bibnamefont {Subramanian}},\
  }\href@noop {} {\bibfield  {journal} {\bibinfo  {journal} {Journal of Fluid
  Mechanics}\ }\textbf {\bibinfo {volume} {890}} (\bibinfo {year}
  {2020}{\natexlab{a}})}\BibitemShut {NoStop}%
\bibitem [{\citenamefont {Ryabov}\ \emph {et~al.}(2021)\citenamefont {Ryabov},
  \citenamefont {Kerimoglu}, \citenamefont {Litchman}, \citenamefont {Olenina},
  \citenamefont {Roselli}, \citenamefont {Basset}, \citenamefont {Stanca},\
  and\ \citenamefont {Blasius}}]{Ryabov_2021}%
  \BibitemOpen
  \bibfield  {author} {\bibinfo {author} {\bibfnamefont {A.}~\bibnamefont
  {Ryabov}}, \bibinfo {author} {\bibfnamefont {O.}~\bibnamefont {Kerimoglu}},
  \bibinfo {author} {\bibfnamefont {E.}~\bibnamefont {Litchman}}, \bibinfo
  {author} {\bibfnamefont {I.}~\bibnamefont {Olenina}}, \bibinfo {author}
  {\bibfnamefont {L.}~\bibnamefont {Roselli}}, \bibinfo {author} {\bibfnamefont
  {A.}~\bibnamefont {Basset}}, \bibinfo {author} {\bibfnamefont
  {E.}~\bibnamefont {Stanca}}, \ and\ \bibinfo {author} {\bibfnamefont
  {B.}~\bibnamefont {Blasius}},\ }\href@noop {} {\bibfield  {journal} {\bibinfo
   {journal} {Ecology Lett.}\ }\textbf {\bibinfo {volume} {24}},\ \bibinfo
  {pages} {847} (\bibinfo {year} {2021})}\BibitemShut {NoStop}%
\bibitem [{\citenamefont {Borgnino}\ \emph {et~al.}(2022)\citenamefont
  {Borgnino}, \citenamefont {Boffetta}, \citenamefont {Cencini}, \citenamefont
  {DeLillo},\ and\ \citenamefont {Gustavsson}}]{Cencini_2022}%
  \BibitemOpen
  \bibfield  {author} {\bibinfo {author} {\bibfnamefont {M.}~\bibnamefont
  {Borgnino}}, \bibinfo {author} {\bibfnamefont {G.}~\bibnamefont {Boffetta}},
  \bibinfo {author} {\bibfnamefont {M.}~\bibnamefont {Cencini}}, \bibinfo
  {author} {\bibfnamefont {F.}~\bibnamefont {DeLillo}}, \ and\ \bibinfo
  {author} {\bibfnamefont {K.}~\bibnamefont {Gustavsson}},\ }\href@noop {}
  {\bibfield  {journal} {\bibinfo  {journal} {Phys. Rev. Fluids}\ }\textbf
  {\bibinfo {volume} {7}},\ \bibinfo {pages} {074603} (\bibinfo {year}
  {2022})}\BibitemShut {NoStop}%
\bibitem [{Note2()}]{Note2}%
  \BibitemOpen
  \bibinfo {note} {An anisotropic passive solute\protect \tmspace +\thinmuskip
  {.1667em}(Brownian fiber) is predicted to exhibit high-shear trapping\cite
  {NitscheHinch_1997}. The predicted inhomogeneity is weak, however, and has
  not been observed in experiments.}\BibitemShut {Stop}%
\bibitem [{Note3()}]{Note3}%
  \BibitemOpen
  \bibinfo {note} {The multiple time scales framework used here assumes the
  swimmer mean free path to be negligibly small, leading to the swimmer
  concentration asymptoting to a Dirac delta function, for $Pe_r \rightarrow
  \infty $, in the centerline-collapse regime,. In practice, swimmers will
  localize in a near-center `Knudsen' layer of a finite thickness.}\BibitemShut
  {Stop}%
\bibitem [{\citenamefont {Subramanian}\ and\ \citenamefont
  {Koch}(2009)}]{SubKoch_2009}%
  \BibitemOpen
  \bibfield  {author} {\bibinfo {author} {\bibfnamefont {G.}~\bibnamefont
  {Subramanian}}\ and\ \bibinfo {author} {\bibfnamefont {D.}~\bibnamefont
  {Koch}},\ }\href@noop {} {\bibfield  {journal} {\bibinfo  {journal} {J. Fluid
  Mech.}\ }\textbf {\bibinfo {volume} {632}},\ \bibinfo {pages} {359} (\bibinfo
  {year} {2009})}\BibitemShut {NoStop}%
\bibitem [{\citenamefont {Saintillan}\ and\ \citenamefont
  {Shelley}(2015)}]{SS_2015}%
  \BibitemOpen
  \bibfield  {author} {\bibinfo {author} {\bibfnamefont {D.}~\bibnamefont
  {Saintillan}}\ and\ \bibinfo {author} {\bibfnamefont {M.}~\bibnamefont
  {Shelley}},\ }\enquote {\bibinfo {title} {Theory of active suspensions},}\
  in\ \href@noop {} {\emph {\bibinfo {booktitle} {Complex Fluids in Biological
  Systems: Experiment, Theory and Computation}}}\ (\bibinfo  {publisher}
  {Springer, New York, NY},\ \bibinfo {year} {2015})\ pp.\ \bibinfo {pages}
  {319--355}\BibitemShut {NoStop}%
\bibitem [{\citenamefont {Vennamneni}\ \emph
  {et~al.}(2020{\natexlab{b}})\citenamefont {Vennamneni}, \citenamefont
  {Garg},\ and\ \citenamefont {Subramanian}}]{ConcBanding_2021}%
  \BibitemOpen
  \bibfield  {author} {\bibinfo {author} {\bibfnamefont {L.}~\bibnamefont
  {Vennamneni}}, \bibinfo {author} {\bibfnamefont {P.}~\bibnamefont {Garg}}, \
  and\ \bibinfo {author} {\bibfnamefont {G.}~\bibnamefont {Subramanian}},\
  }\href@noop {} {\bibfield  {journal} {\bibinfo  {journal} {J. Fluid Mech.}\
  }\textbf {\bibinfo {volume} {904}},\ \bibinfo {pages} {A7} (\bibinfo {year}
  {2020}{\natexlab{b}})}\BibitemShut {NoStop}%
\bibitem [{\citenamefont {Jeffery}(1922)}]{Jeffery1922}%
  \BibitemOpen
  \bibfield  {author} {\bibinfo {author} {\bibfnamefont {G.}~\bibnamefont
  {Jeffery}},\ }\href@noop {} {\bibfield  {journal} {\bibinfo  {journal} {Proc.
  Roy. Soc. Lond.}\ }\textbf {\bibinfo {volume} {102}},\ \bibinfo {pages} {161}
  (\bibinfo {year} {1922})}\BibitemShut {NoStop}%
\bibitem [{\citenamefont {Leal}\ and\ \citenamefont
  {Hinch}(1971)}]{HinchLeal1971}%
  \BibitemOpen
  \bibfield  {author} {\bibinfo {author} {\bibfnamefont {L.}~\bibnamefont
  {Leal}}\ and\ \bibinfo {author} {\bibfnamefont {E.}~\bibnamefont {Hinch}},\
  }\href@noop {} {\bibfield  {journal} {\bibinfo  {journal} {Journal of Fluid
  Mechanics}\ }\textbf {\bibinfo {volume} {46}},\ \bibinfo {pages} {685}
  (\bibinfo {year} {1971})}\BibitemShut {NoStop}%
\bibitem [{Note4()}]{Note4}%
  \BibitemOpen
  \bibinfo {note} {These are nominal estimates for $Pe_r \sim O(1)$; actual
  estimates involve additional $Pe_r$-dependent factors in both high and
  low-shear trapping regimes\cite {vennamneni2020shear}.}\BibitemShut {Stop}%
\bibitem [{\citenamefont {Subramanian}\ and\ \citenamefont
  {Brady}(2004)}]{SubBrady_2004}%
  \BibitemOpen
  \bibfield  {author} {\bibinfo {author} {\bibfnamefont {G.}~\bibnamefont
  {Subramanian}}\ and\ \bibinfo {author} {\bibfnamefont {J.}~\bibnamefont
  {Brady}},\ }\href@noop {} {\bibfield  {journal} {\bibinfo  {journal} {Physica
  A}\ }\textbf {\bibinfo {volume} {334}},\ \bibinfo {pages} {343} (\bibinfo
  {year} {2004})}\BibitemShut {NoStop}%
\bibitem [{SM()}]{SM}%
  \BibitemOpen
  \href@noop {} {\bibinfo  {journal} {See Supplementary information}\
  }\BibitemShut {NoStop}%
\bibitem [{Note5()}]{Note5}%
  \BibitemOpen
\bibfield  {journal} {  }\bibinfo {note} {The uniform state is the result of
  neglecting direct interactions of the swimmers, hydrodynamic or otherwise,
  with boundaries; such interactions lead to localized wall layers which play a
  significant role for narrow channels corresponding to finite $\epsilon $\cite
  {Gompper1,Gompper2,Underhill2015}.}\BibitemShut {Stop}%
\bibitem [{\citenamefont {Dehkharghani}\ \emph {et~al.}(2019)\citenamefont
  {Dehkharghani}, \citenamefont {Waisbord}, \citenamefont {Dunkel},\ and\
  \citenamefont {Guasto}}]{GuastoPNAS2019}%
  \BibitemOpen
  \bibfield  {author} {\bibinfo {author} {\bibfnamefont {A.}~\bibnamefont
  {Dehkharghani}}, \bibinfo {author} {\bibfnamefont {N.}~\bibnamefont
  {Waisbord}}, \bibinfo {author} {\bibfnamefont {J.}~\bibnamefont {Dunkel}}, \
  and\ \bibinfo {author} {\bibfnamefont {J.}~\bibnamefont {Guasto}},\
  }\href@noop {} {\bibfield  {journal} {\bibinfo  {journal} {Proc. Natl. Acad.
  Sci.}\ }\textbf {\bibinfo {volume} {116}},\ \bibinfo {pages} {11119}
  (\bibinfo {year} {2019})}\BibitemShut {NoStop}%
\bibitem [{\citenamefont {Haynes}\ and\ \citenamefont
  {Vanneste}(2014)}]{Haynes_2014}%
  \BibitemOpen
  \bibfield  {author} {\bibinfo {author} {\bibfnamefont {P.}~\bibnamefont
  {Haynes}}\ and\ \bibinfo {author} {\bibfnamefont {J.}~\bibnamefont
  {Vanneste}},\ }\href@noop {} {\bibfield  {journal} {\bibinfo  {journal} {J.
  Fluid Mech.}\ }\textbf {\bibinfo {volume} {745}},\ \bibinfo {pages} {32}
  (\bibinfo {year} {2014})}\BibitemShut {NoStop}%
\bibitem [{\citenamefont {Sreenivasan}(1997)}]{KRS_1997}%
  \BibitemOpen
  \bibfield  {author} {\bibinfo {author} {\bibfnamefont {K.}~\bibnamefont
  {Sreenivasan}},\ }\href@noop {} {\bibfield  {journal} {\bibinfo  {journal}
  {Ann. Rev. Fluid Mech.}\ }\textbf {\bibinfo {volume} {29}},\ \bibinfo {pages}
  {435} (\bibinfo {year} {1997})}\BibitemShut {NoStop}%
\bibitem [{\citenamefont {Hill}\ \emph {et~al.}(2007)\citenamefont {Hill},
  \citenamefont {Ozge}, \citenamefont {McMurry},\ and\ \citenamefont
  {Koser}}]{Hill_2007}%
  \BibitemOpen
  \bibfield  {author} {\bibinfo {author} {\bibfnamefont {J.}~\bibnamefont
  {Hill}}, \bibinfo {author} {\bibfnamefont {K.}~\bibnamefont {Ozge}}, \bibinfo
  {author} {\bibfnamefont {J.}~\bibnamefont {McMurry}}, \ and\ \bibinfo
  {author} {\bibfnamefont {H.}~\bibnamefont {Koser}},\ }\href@noop {}
  {\bibfield  {journal} {\bibinfo  {journal} {Phys. Rev. Lett.}\ }\textbf
  {\bibinfo {volume} {98}},\ \bibinfo {pages} {068101} (\bibinfo {year}
  {2007})}\BibitemShut {NoStop}%
\bibitem [{\citenamefont {Costanzo}\ \emph {et~al.}(2012)\citenamefont
  {Costanzo}, \citenamefont {Leonardo}, \citenamefont {Ruocco},\ and\
  \citenamefont {Angelani}}]{Costanzo_2012}%
  \BibitemOpen
  \bibfield  {author} {\bibinfo {author} {\bibfnamefont {A.}~\bibnamefont
  {Costanzo}}, \bibinfo {author} {\bibfnamefont {R.}~\bibnamefont {Leonardo}},
  \bibinfo {author} {\bibfnamefont {G.}~\bibnamefont {Ruocco}}, \ and\ \bibinfo
  {author} {\bibfnamefont {L.}~\bibnamefont {Angelani}},\ }\href@noop {}
  {\bibfield  {journal} {\bibinfo  {journal} {J. Phys.: Condes. Matter}\
  }\textbf {\bibinfo {volume} {24}},\ \bibinfo {pages} {065101} (\bibinfo
  {year} {2012})}\BibitemShut {NoStop}%
\bibitem [{\citenamefont {Kaya}\ and\ \citenamefont {Koser}(2012)}]{Kaya_2012}%
  \BibitemOpen
  \bibfield  {author} {\bibinfo {author} {\bibfnamefont {T.}~\bibnamefont
  {Kaya}}\ and\ \bibinfo {author} {\bibfnamefont {H.}~\bibnamefont {Koser}},\
  }\href@noop {} {\bibfield  {journal} {\bibinfo  {journal} {Biophys. J.}\
  }\textbf {\bibinfo {volume} {102}},\ \bibinfo {pages} {1514} (\bibinfo {year}
  {2012})}\BibitemShut {NoStop}%
\bibitem [{\citenamefont {Figueroa-Morales}\ \emph {et~al.}(2015)\citenamefont
  {Figueroa-Morales}, \citenamefont {Mino}, \citenamefont {Rivera},
  \citenamefont {Caballero}, \citenamefont {Clement}, \citenamefont
  {Altshuler},\ and\ \citenamefont {Lindner}}]{Lindner_2015}%
  \BibitemOpen
  \bibfield  {author} {\bibinfo {author} {\bibfnamefont {N.}~\bibnamefont
  {Figueroa-Morales}}, \bibinfo {author} {\bibfnamefont {G.}~\bibnamefont
  {Mino}}, \bibinfo {author} {\bibfnamefont {A.}~\bibnamefont {Rivera}},
  \bibinfo {author} {\bibfnamefont {R.}~\bibnamefont {Caballero}}, \bibinfo
  {author} {\bibfnamefont {E.}~\bibnamefont {Clement}}, \bibinfo {author}
  {\bibfnamefont {E.}~\bibnamefont {Altshuler}}, \ and\ \bibinfo {author}
  {\bibfnamefont {A.}~\bibnamefont {Lindner}},\ }\href@noop {} {\bibfield
  {journal} {\bibinfo  {journal} {Soft Matter}\ }\textbf {\bibinfo {volume}
  {11}},\ \bibinfo {pages} {6284} (\bibinfo {year} {2015})}\BibitemShut
  {NoStop}%
\bibitem [{\citenamefont {Vennamneni}\ and\ \citenamefont
  {Subramanian}(tted)}]{ProlateOblateman}%
  \BibitemOpen
  \bibfield  {author} {\bibinfo {author} {\bibfnamefont {L.}~\bibnamefont
  {Vennamneni}}\ and\ \bibinfo {author} {\bibfnamefont {G.}~\bibnamefont
  {Subramanian}},\ }\href@noop {} {\bibfield  {journal} {\bibinfo  {journal}
  {J. Fluid Mech.}\ } (\bibinfo {year} {to be submitted})}\BibitemShut
  {NoStop}%
\bibitem [{\citenamefont {Saintillan}\ and\ \citenamefont
  {Shelley}(2007)}]{SS2007}%
  \BibitemOpen
  \bibfield  {author} {\bibinfo {author} {\bibfnamefont {D.}~\bibnamefont
  {Saintillan}}\ and\ \bibinfo {author} {\bibfnamefont {M.}~\bibnamefont
  {Shelley}},\ }\href@noop {} {\bibfield  {journal} {\bibinfo  {journal} {Phys.
  Rev. Lett.}\ }\textbf {\bibinfo {volume} {99}},\ \bibinfo {pages} {058102}
  (\bibinfo {year} {2007})}\BibitemShut {NoStop}%
\bibitem [{\citenamefont {Saintillan}\ and\ \citenamefont
  {Shelley}(2008)}]{SS2008}%
  \BibitemOpen
  \bibfield  {author} {\bibinfo {author} {\bibfnamefont {D.}~\bibnamefont
  {Saintillan}}\ and\ \bibinfo {author} {\bibfnamefont {M.}~\bibnamefont
  {Shelley}},\ }\href@noop {} {\bibfield  {journal} {\bibinfo  {journal} {Phys.
  Fluids}\ }\textbf {\bibinfo {volume} {20}},\ \bibinfo {pages} {123304}
  (\bibinfo {year} {2008})}\BibitemShut {NoStop}%
\bibitem [{\citenamefont {Koch}\ and\ \citenamefont
  {Subramanian}(2011)}]{KochSub_2011}%
  \BibitemOpen
  \bibfield  {author} {\bibinfo {author} {\bibfnamefont {D.}~\bibnamefont
  {Koch}}\ and\ \bibinfo {author} {\bibfnamefont {G.}~\bibnamefont
  {Subramanian}},\ }\href@noop {} {\bibfield  {journal} {\bibinfo  {journal}
  {Ann. Rev. Fluid Mech.}\ }\textbf {\bibinfo {volume} {43}},\ \bibinfo {pages}
  {637} (\bibinfo {year} {2011})}\BibitemShut {NoStop}%
\bibitem [{\citenamefont {Krishnamurthy}\ and\ \citenamefont
  {Subramanian}(2015)}]{Deepak_2015}%
  \BibitemOpen
  \bibfield  {author} {\bibinfo {author} {\bibfnamefont {D.}~\bibnamefont
  {Krishnamurthy}}\ and\ \bibinfo {author} {\bibfnamefont {G.}~\bibnamefont
  {Subramanian}},\ }\href@noop {} {\bibfield  {journal} {\bibinfo  {journal}
  {J. Fluid Mech.}\ }\textbf {\bibinfo {volume} {781}},\ \bibinfo {pages} {422}
  (\bibinfo {year} {2015})}\BibitemShut {NoStop}%
\bibitem [{\citenamefont {Martinez}\ \emph {et~al.}(2020)\citenamefont
  {Martinez}, \citenamefont {Clement}, \citenamefont {Arit},\ and\
  \citenamefont {Poon}}]{Martinez_2020}%
  \BibitemOpen
  \bibfield  {author} {\bibinfo {author} {\bibfnamefont {V.}~\bibnamefont
  {Martinez}}, \bibinfo {author} {\bibfnamefont {E.}~\bibnamefont {Clement}},
  \bibinfo {author} {\bibfnamefont {J.}~\bibnamefont {Arit}}, \ and\ \bibinfo
  {author} {\bibfnamefont {W.}~\bibnamefont {Poon}},\ }\href@noop {} {\bibfield
   {journal} {\bibinfo  {journal} {Proc. Natl. Acad. Sci.}\ }\textbf {\bibinfo
  {volume} {117}},\ \bibinfo {pages} {2326} (\bibinfo {year}
  {2020})}\BibitemShut {NoStop}%
\bibitem [{\citenamefont {Taylor}(1954)}]{Taylor_1954}%
  \BibitemOpen
  \bibfield  {author} {\bibinfo {author} {\bibfnamefont {G.}~\bibnamefont
  {Taylor}},\ }\href@noop {} {\bibfield  {journal} {\bibinfo  {journal} {Prof.
  R. Soc. Lond.}\ }\textbf {\bibinfo {volume} {A223}},\ \bibinfo {pages} {446}
  (\bibinfo {year} {1954})}\BibitemShut {NoStop}%
\bibitem [{\citenamefont {Foister}\ and\ \citenamefont {Van
  De~Ven}(1980)}]{Foister_1980}%
  \BibitemOpen
  \bibfield  {author} {\bibinfo {author} {\bibfnamefont {R.}~\bibnamefont
  {Foister}}\ and\ \bibinfo {author} {\bibfnamefont {T.}~\bibnamefont {Van
  De~Ven}},\ }\href@noop {} {\bibfield  {journal} {\bibinfo  {journal} {Journal
  of Fluid Mechanics}\ }\textbf {\bibinfo {volume} {96}},\ \bibinfo {pages}
  {105} (\bibinfo {year} {1980})}\BibitemShut {NoStop}%
\bibitem [{\citenamefont {Nitsche}\ and\ \citenamefont
  {Hinch}(1997)}]{NitscheHinch_1997}%
  \BibitemOpen
  \bibfield  {author} {\bibinfo {author} {\bibfnamefont {L.}~\bibnamefont
  {Nitsche}}\ and\ \bibinfo {author} {\bibfnamefont {E.}~\bibnamefont
  {Hinch}},\ }\href@noop {} {\bibfield  {journal} {\bibinfo  {journal} {J.
  Fluid Mech.}\ }\textbf {\bibinfo {volume} {332}},\ \bibinfo {pages} {1}
  (\bibinfo {year} {1997})}\BibitemShut {NoStop}%
\bibitem [{\citenamefont {Elgeti}\ and\ \citenamefont
  {Gompper}(2009)}]{Gompper1}%
  \BibitemOpen
  \bibfield  {author} {\bibinfo {author} {\bibfnamefont {J.}~\bibnamefont
  {Elgeti}}\ and\ \bibinfo {author} {\bibfnamefont {G.}~\bibnamefont
  {Gompper}},\ }\href@noop {} {\bibfield  {journal} {\bibinfo  {journal} {Eur.
  Phys. Lett.}\ }\textbf {\bibinfo {volume} {85}},\ \bibinfo {pages} {38002}
  (\bibinfo {year} {2009})}\BibitemShut {NoStop}%
\bibitem [{\citenamefont {Elgeti}\ and\ \citenamefont
  {Gompper}(2013)}]{Gompper2}%
  \BibitemOpen
  \bibfield  {author} {\bibinfo {author} {\bibfnamefont {J.}~\bibnamefont
  {Elgeti}}\ and\ \bibinfo {author} {\bibfnamefont {G.}~\bibnamefont
  {Gompper}},\ }\href@noop {} {\bibfield  {journal} {\bibinfo  {journal} {Eur.
  Phys. Lett.}\ }\textbf {\bibinfo {volume} {101}},\ \bibinfo {pages} {48003}
  (\bibinfo {year} {2013})}\BibitemShut {NoStop}%
\bibitem [{\citenamefont {Chilukuri}\ \emph {et~al.}(2015)\citenamefont
  {Chilukuri}, \citenamefont {Collins},\ and\ \citenamefont
  {Underhill}}]{Underhill2015}%
  \BibitemOpen
  \bibfield  {author} {\bibinfo {author} {\bibfnamefont {S.}~\bibnamefont
  {Chilukuri}}, \bibinfo {author} {\bibfnamefont {C.}~\bibnamefont {Collins}},
  \ and\ \bibinfo {author} {\bibfnamefont {P.}~\bibnamefont {Underhill}},\
  }\href@noop {} {\bibfield  {journal} {\bibinfo  {journal} {Phys. Fluids}\
  }\textbf {\bibinfo {volume} {27}},\ \bibinfo {pages} {031902} (\bibinfo
  {year} {2015})}\BibitemShut {NoStop}%
\end{thebibliography}%

\end{document}